\documentclass[prl, aps, twocolumn, superscriptaddress,noshowpacs]{revtex4}
\usepackage[english]{babel} 
\usepackage[english]{layout}
\usepackage{amsmath,amsfonts,amssymb}
\usepackage{graphicx}
\usepackage{mathrsfs}
\usepackage{float}
\usepackage{color}
\usepackage{bm}
\usepackage{braket}
\usepackage{version}
\usepackage{array,multirow,makecell}
\usepackage[T1]{fontenc}

\begin{document}

\title{Weyl singularities in polaritonic multi-terminal Josephson junctions}
\author{I. Septembre}
\affiliation{Universit\'e Clermont Auvergne, Clermont Auvergne INP, CNRS, Institut Pascal, F-63000 Clermont-Ferrand, France}
\author{J. S. Meyer}
\affiliation{Univ. Grenoble Alpes, CEA, Grenoble INP, IRIG, Pheliqs, 38000 Grenoble, France}
\author{D. D. Solnyshkov}
\affiliation{Universit\'e Clermont Auvergne, Clermont Auvergne INP, CNRS, Institut Pascal, F-63000 Clermont-Ferrand, France}
\affiliation{Institut Universitaire de France (IUF), 75231 Paris, France}
\author{G. Malpuech}
\affiliation{Universit\'e Clermont Auvergne, Clermont Auvergne INP, CNRS, Institut Pascal, F-63000 Clermont-Ferrand, France}

\begin{abstract}
We study theoretically analog multi-terminal Josephson junctions formed by gapped superfluids created upon resonant pumping of cavity exciton-polaritons. We study the $p$-like bands of a 5-terminal junction in the 4D parameter space created by the superfluid phases acting as quasi-momenta. We find 4/6 Weyl points in 3D subspaces with preserved/broken time-reversal symmetry. We link the real space topology (vortices) to the parameter space one (Weyl points). We derive an effective Hamiltonian encoding the creation, motion, and annihilation of Weyl nodes in 4D. Our work paves the way to the study of exotic topological phases in a platform allowing direct measurement of eigenstates and band topology.
\end{abstract}

\maketitle
\emph{Introduction.--}
Topological singularities are singularities of both eigenvalues and eigenstates carrying a topological charge. Dirac points are 2D topological point singularities. They transform into pairs of exceptional points connected by a Fermi arc when adding non-Hermiticity~\cite{Voigt1902,Richter2019,krol2022annihilation}. In 3D parameter spaces, Weyl points (WPs) are Hermitian point degeneracies~\cite{lu2015experimental,noh2017experimental} (different from 3D Dirac points~\cite{young2012dirac,armitage2018weyl}). They are robust because any Hermitian perturbation only moves the WPs in the parameter space, whereas it can destroy Dirac points. WPs resemble exceptional points because they come in pairs connected by Fermi arcs~\cite{noh2017experimental,zhou2018observation}. The only way to annihilate both Weyl and exceptional points is to make points of opposite charge meet~\cite{krol2022annihilation,liu2021observation}. WPs can appear when time-reversal (TR) symmetry and/or inversion symmetry is broken. If TR symmetry is preserved, they come in multiples of 4; they come in multiples of 2 otherwise~\cite{nielsen1981absence}. %\cite{liu2018giant,borisenko2019time,wang2019photonic}  
Furthermore, additional symmetries make appear nodal lines rather than points~\cite{yu2015topological,kim2015dirac,xie2015new,sun2018conversion,gao2018experimental,xia2019observation}. They are line singularities in 3D linked giving drumhead surface states~\cite{bian2016drumhead,chan20163,belopolski2019discovery}.
Topological bands and states benefit from outstanding properties such as (in 2D) one-way edge propagation~\cite{haldane2008possible,raghu2008analogs,wang2008reflection,wang2009observation,klembt2018exciton,barik2018topological,PhysRevLett.129.066802}, used in topological lasers~\cite{solnyshkov2016kibble,st2017lasing,bandres2018topological,gong2020topological,zeng2020electrically}.

Topological singularities described previously lie in a parameter space which is usually the reciprocal space, because it naturally comes as the matching of the real space.  
However, some parameters that are easily tunable experimentally can form additional dimensions of the parameter space. 
This enriches the exploration of topological phase transitions. Such systems are called \textit{synthetic topological matter}. The fantastic freedom they offer is an enthralling playground for physicists~\cite{PhysRevLett.112.043001,goldman2016topological,ozawa2019topological,chen2021creating}. It enables to investigate physics beyond 3~dimensions~\cite{PhysRevLett.108.133001,PhysRevLett.115.195303,lohse2018exploring} as well as strongly correlated phenomena~\cite{junemann2017exploring,jian2018interacting,elben2020many}. Topological photonics~\cite{yang2020mode,lustig2021topological}, and especially polaritonics~\cite{pickup2020synthetic}, can take advantage of synthetic topological matter, notably because the latter gives an experimental access to the eigenstates and quantum geometry of topological systems~\cite{gianfrate2020measurement,Liao2021}.

Andreev reflection occurs at the interface between a superconductor and a non-superconducting material~\cite{andreev1964thermal}. An incoming electron undergoes an anomalous reflection, becoming a hole excitation with reversed wavevector, charge, and spin. Usual Josephson junctions~\cite{josephson1962possible} contain two interfaces between a non-superconducting material (insulator, semiconductor, or metal) and a superconductor. Such junctions host Andreev bound states whose energy depends on the phase difference between the superconductors. This dependence can be described in terms of synthetic bands, where the 1D parameter space is here given by the phase difference~\cite{skoldberg2008spectrum,kotetes2019synthetic}. If the superconductors are topological, those synthetic bands can be topological when the energies cross at the Fermi energy, which forms a topological singularity. In this case, the junction is known to host Majorana fermions~\cite{majorana1937teoria}, which are very promising for quantum computing~\cite{aasen2016milestones}.
Multi-terminal Josephson junctions, where more than two superconductor wires are connected~\cite{riwar2016multi}, is now a well-developed research area~\cite{yokoyama2015singularities,PhysRevLett.119.136807,eriksson2017topological,xie2017topological,deb2018josephson,houzet2019majorana,xie2019topological,gavensky2019topological,Klees2020,klees2021ground,weisbrich2021second}. The corresponding synthetic bands demonstrate non-trivial topology (in arbitrary large dimensions) even with trivial superconductors.

Andreev reflection has been studied theoretically in bosonic systems~\cite{Zapata2009}. In cavity exciton-polaritons, an analog superconductor can be created upon resonant driving. The driving opens a gap in the energy spectrum of the pumped modes, creating a ``gapped superfluid''~\cite{carusotto2004probing,ciuti2005quantum,Carusotto2013,kavokin2017microcavities}. It is possible to create analog Josephson junctions by pumping two (or more) regions with different pump lasers (with well-defined phases), the superfluids being separated by a common non-superfluid region~\cite{goblot2016phase,Koniakhin2019,claude2020taming}. In~\cite{septembre2021parametric}, the existence of Andreev bound state analogs in the normal region between two gapped superfluids has been demonstrated, while the 1D bands parameterized by the superfluid phase difference were found to be separated by a topological gap. 

In this work, we theoretically study a multi-terminal polaritonic Josephson junction, where a pentagonal normal region is connected to 5 gapped superfluids. Such a junction hosts Andreev-like bound states which form 4D synthetic bands. We study the pair of bands corresponding to the $p$-like confined states of the pentagonal trap. 
We study numerically a 3D subspace and observe 4 WPs. These WPs enable topological transitions in a 2D subspace between distinct topological phases with different Chern numbers. 
We model the 2 $p$-like bands by an effective $2\times 2$ Hamiltonian and explore the full 4D parameter space, leading to additional features such as 3D subspaces with broken TR symmetry.
This photonic system has the advantage of allowing the access to the eigenstates, making possible the direct measurement of the Berry curvature and the detailed study of the WPs.
Our results may allow to finally observe Weyl singularities in multi-terminal Josephson junctions and pave the way to synthetic topological matter in arbitrary large dimensions in polaritonics.

\begin{figure}[tbp]
    \centering
    \includegraphics[width=0.99\linewidth]{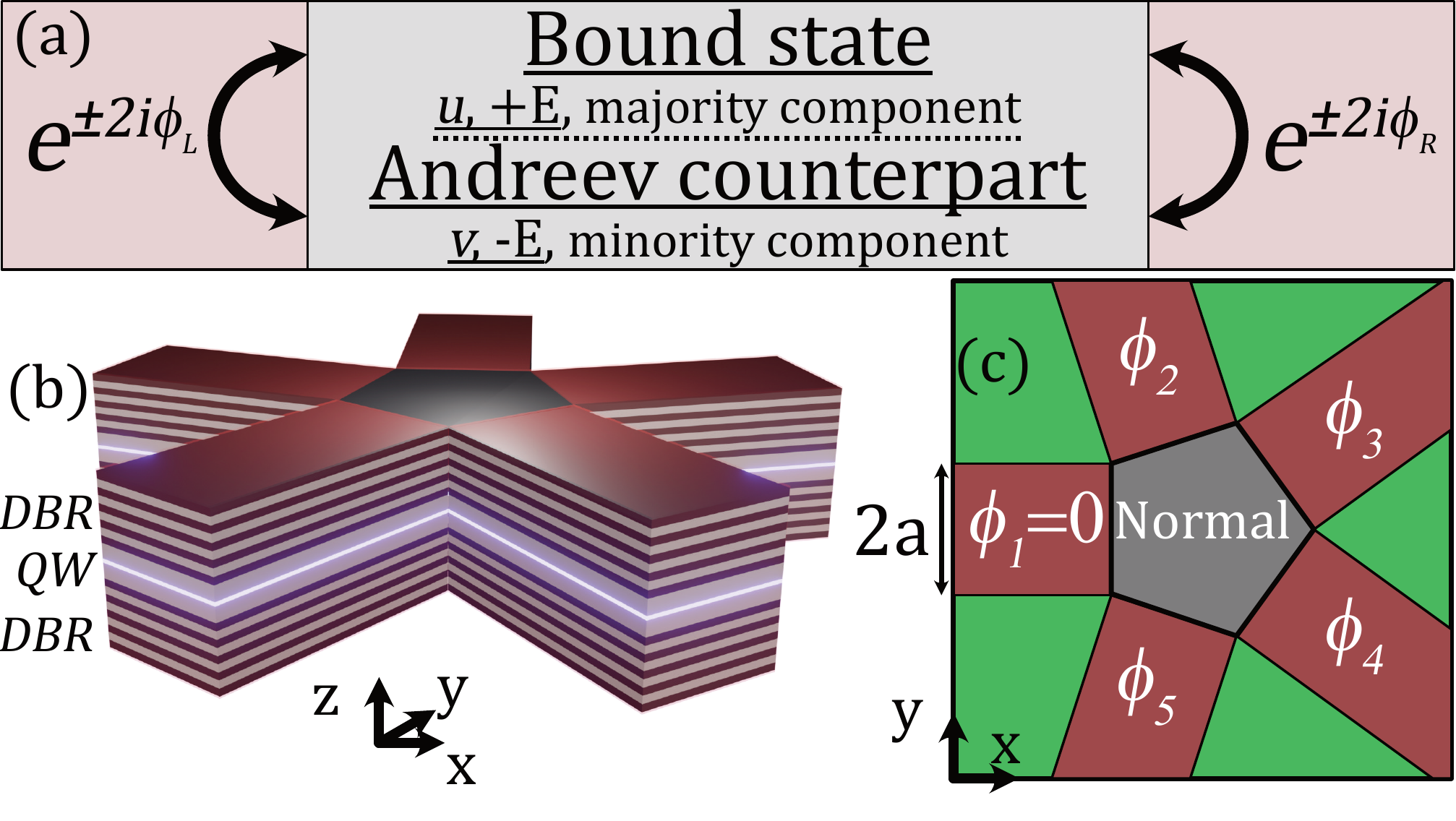}
    \caption{(a) 1D superfluid-normal-superfluid junction hosting bound states and their Andreev counterparts, linked by the phases of the left/right superfluid $\phi_{L,R}$. (b) Sketch of the analog multi-terminal Josephson junction (here 5-terminal) organized in a microcavity with a quantum well (QW)  and distributed Bragg reflectors (DBR) on a substrate. (c) Top view of the junction. The superfluid density (brown) with 5 different phases $\phi_i$ and the potential $\mathcal{V}(\textbf{r})$ (green) profiles. In the normal region, pumping and potential are absent.}
    \label{fig1}
\end{figure}

\emph{Model.--}
We consider a strongly-coupled planar microcavity hosting exciton-polaritons~\cite{Carusotto2013,goblot2016phase,kavokin2017microcavities}. The external drive $P(\textbf{r})$ is composed of $\mathscr{N}$ regions, each pumping a given area with an homogeneous amplitude $P_i=e^{-i (\omega_p t+\phi'_i)}$ ($i\in [1,\mathscr{N}]$), where $E_p=\hbar \omega_p$ is the pump detuning and $\phi'_i$ the phase of the $i$-th region of the pump, determining the phase of the corresponding superfluid $\phi_i$~\cite{Koniakhin2019,suppl}. This forms an analog $\mathscr{N}$-terminal Josephson junction as depicted in Fig.~\ref{fig1}. The wave function $\psi$ of the coherent pumped mode is described by the driven-dissipative Gross-Pitaevskii equation: 
\begin{equation}\label{GP}
    i\hbar \frac{\partial\psi}{\partial t}=\left[ -\frac{\hbar^2}{2m}\nabla^2 -i\gamma +\alpha(\mathbf{r}) |\psi|^2 +\mathcal{V}(\mathbf{r})\right] \psi +P(\mathbf{r}),
\end{equation}
where $m$ is the exciton-polaritons effective mass, $\gamma$ the decay (further taken equal to zero, it only adds a global imaginary part), $\alpha>0$ describes the repulsive interactions in the pumped areas, and  $\mathcal{V}(\textbf{r})$ is a step-like potential (in green in Fig.~\ref{fig1}(c)) accounting for the etched pattern. We assume that the stationary wave function $\psi_s$ is zero in all regions without pumping. 
In the $\mathscr{N}$ pumped areas, the stationary wave function is given by the solution of the Gross-Pitaevskii equation for a spatially homogeneous system and reads:
$\psi_{s,i}=\sqrt{n}e^{i\phi_i}$ where $n$ is the superfluid density and $\phi_i$ its phase. The spectrum of the superfluid weak excitations in these areas contains a gap $\Delta$ centered around the pump detuning $E_p$:
\begin{equation}
    \Delta=\sqrt{\left( 3\alpha n - E_p\right)\left( \alpha n - E_p\right)}
\end{equation}

We consider weak excitations of the full system (the normal and superfluid areas), looking for the solutions of the following shape:
\begin{equation}\label{wpwf}
    \psi(\mathbf{r},t)=e^{-i\omega_p t}\left(\psi_s(\mathbf{r})+ u(\mathbf{r})e^{-i\omega t}+v^*(\mathbf{r}) e^{i\omega^* t}\right),
\end{equation}
where $u(\mathbf{r}),~v(\mathbf{r})$ are the Bogoliubov coefficients. We will consider energies lying in the superfluid gap, so that these coefficients describe the profile of propagative states in the normal region and evanescent states in the superfluids, which gives precisely a bound state.
This wave function, inserted in Eq.~\eqref{GP}, gives the Bogoliubov-de Gennes equations:
\begin{equation}\label{BdG0}
\begin{pmatrix}
 \mathscr{L} && \alpha \psi_s^2(\mathbf{r}) \\
 -\alpha \psi_s^{*2}(\mathbf{r})  && -\mathscr{L^*}
\end{pmatrix}\begin{pmatrix}
u(\mathbf{r})\\v(\mathbf{r})
\end{pmatrix}=\hbar\omega\begin{pmatrix}
u(\mathbf{r})\\v(\mathbf{r})
\end{pmatrix},
\end{equation}
where $\mathscr{L}=\left(\epsilon_\mathbf{k}-E_p+2\alpha n + \mathcal{V}(\mathbf{r}) \right)$ with $\epsilon_\mathbf{k}=\hbar ^2 \mathbf{k} ^2/2m$.

The case $\mathscr{N}=2$ is studied in~\cite{septembre2021parametric};
$\mathscr{N}=2$ corresponds to an analog Josephson junction with 2 terminals, containing two normal/superfluid interfaces (phases $\phi_{L,R}$), as depicted in Fig.~\ref{fig1}(a). A wave incident at each interface undergoes both specular reflection (same energy, reversed wavevector) and Andreev reflection (opposite energy with respect to the pump $E_p$, reversed wavevector), the latter being accompanied by a phase shift $e^{\pm 2i \phi_{L,R}}$. Considering the complete scattering process, Andreev bound states can be found. These states are composed of two parts of different nature. First, the majority component of profile $u(\bm{r})$ and energy $E=\hbar\omega$ is qualitatively the state coming from the quantum confinement provided by the interactions in the pumped regions. It exists without Andreev process and corresponds to the state found with a diagonal matrix in Eq.~\eqref{BdG0}.%, without coupling between $u$ and $v$. 
The second part is the minority component of profile $v(\bm{r})$ and energy $-E$, which is not (in general) an eigenstate of the quantum well formed by the junction. It appears only because of the Andreev reflections at the interfaces and is therefore very sensitive to the phase difference between the two superfluids. We call it the minority component, or Andreev counterpart of the bound state. The phase difference between the two superfluids is a parameter that can be tuned experimentally. Both the majority component and its Andreev counterpart form synthetic bands in the 1D space with a dispersion $E(a,\phi)=E_0(a)\cos(2(\phi_R-\phi_L))$, where $E_0(a)$ evolves as in a usual quantum well. In 1D, only the Andreev counterpart carries topology (non-zero Zak phase~\cite{Zak1989}, a topological invariant for bandgaps~\cite{Delplace2011}, as in the Su-Schrieffer-Heeger model~\cite{Su1979}). Topological transitions occur between bands of different nature (majority/minority). They interact through the off-diagonal terms of the anti-Hermitian matrix~\eqref{BdG0}. Instead of crossing, the bands merge, forming a Fermi arc connecting two exceptional points~\cite{septembre2021parametric}.

In the following, we consider a 2D real space with 5 terminals. This gives access to 4D synthetic bands and to regimes forbidden in 1D, in particular the formation of Weyl singularities.

% ============================ Weyl points ============================ %
\emph{Weyl points.--}
\begin{figure*}[tbp]
    \centering
    \includegraphics[width=0.99\linewidth]{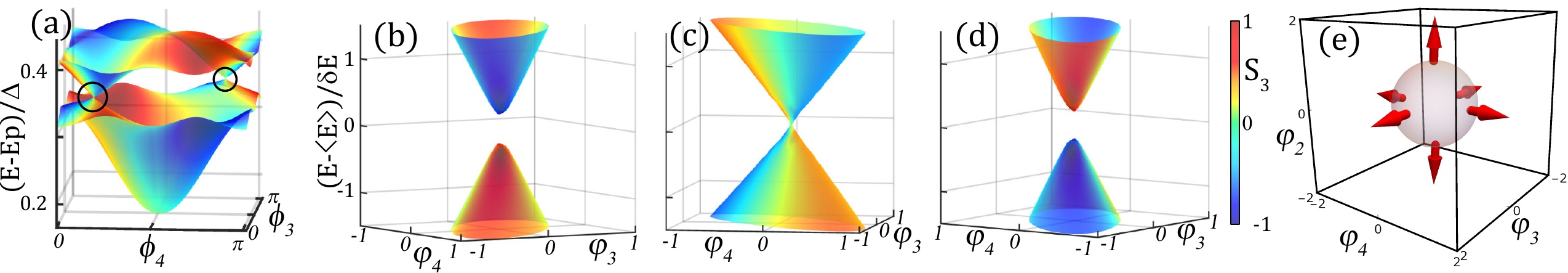}
    \caption{(a) Energy spectrum of the two $\Tilde{p}$ bands for $\phi_2=-\phi_5=\pi/2$. Note the 2 linear crossing points (black circles). (b-d) Energy spectra close to a WP for $\phi_2=-\phi_5=\pi/2+\varepsilon(\pi/20)$ with $\varepsilon=-1$ (b), $\varepsilon=0$ (c), and $\varepsilon=1$ (d). In (a-d), false color shows the pseudo-spin $S_3$. Note the gap closing and opening between (b,c,d) with opposite $S_3$ at the extrema of the bands. (e) Numerically calculated 3D pseudo-spin texture centered around a WP. The sphere represents points at equal distance from the WP. $\bm{\varphi}=(\bm{\phi}-\bm{\phi}^{(0)})/(\pi/20)$ is the reduced parameter space in the vicinity of a WP of coordinates $\bm{\phi}^{(0)}$.}
    \label{fig2}
\end{figure*}
The 5-terminal Josephson junction considered is sketched in Fig.~\ref{fig1}(b). The normal region forms a regular pentagonal prism (a regular pentagon in the $(x,y)$ plane) around which the superfluid regions are regularly placed. Solving numerically Eq.~\eqref{BdG0} using the potentials and pumping terms explained in Fig.~\ref{fig1}(c) allows one to find bound states. We focus on the majority components of the two isolated $\Tilde{p}$ states of the pentagonal potential trap~\cite{suppl}.
The $\Tilde{p}$ states can be decomposed on a ($\Tilde{p}_x$,$\Tilde{p}_y$) basis which naturally leads to three pseudo-spin components $S_{1,2,3}$.
Qualitatively, the components denote linear, diagonal, and circular (vortex) shape of the state, respectively. We represent $(S_2,S_1)$ by arrows and $S_3$ by colors.

We set $\phi_1=0$ as a phase reference and start by studying the 3D TR symmetric subspace $(\phi_2=-\phi_5,\phi_3,\phi_4)$. We choose this configuration because it attributes symmetric roles to $\phi_{3,4}$ and facilitates the numerical study. The 2D bands (versus $\phi_3,\phi_4$) and their $S_3$ components for $\phi_2=-\phi_5=\pi/2$ are shown in Fig.~\ref{fig2}(a). They are degenerate in two points along the diagonal direction $\phi_3=\phi_4$. 
Fig.~\ref{fig2}(c) is a close-up of panel (a) near a linear crossing point. Decreasing (resp. increasing) slightly $\phi_2$, the crossing point disappears and a gap forms, as depicted in Fig.~\ref{fig2}(b) (resp. (d)). A clear change in $S_3$ is observable at the extrema of the two bands (where they are the closest): $S_3=+1$ (red color) for the lower band and $-1$ (blue) for the upper one in Fig.~\ref{fig2}(b) whereas it is the opposite after the band crossing in Fig.~\ref{fig2}(d). This suggests that the gap closure through the WP is indeed a topological transition. 
This can be further demonstrated by computing the 
band Chern number~\cite{berry1984quantal,RevModPhys.82.3045,riwar2016multi}:
\begin{equation}\label{Chern}
    \mathscr{C}=\frac{1}{2\pi}\iint_{S_\phi} \mathscr{B}(\phi_3,\phi_4)\, d\phi_3\,d\phi_4,
\end{equation}
where $\mathscr{B}$ is the Berry curvature and $S_\phi=[0,\pi]^2$ the first synthetic Brillouin zone. 
The gap Chern number (the Chern number of the lower band in this 2-band system~\cite{suppl}) is -1 for Fig.~\ref{fig2}(b) and +1 for Fig.~\ref{fig2}(d), which proves the occurrence of a topological inversion.
Finally, Fig.~\ref{fig2}(e) shows the texture of the effective field %(field parallel to the lowest energy state pseudo-spin) 
on a spherical isoenergetic surface surrounding the WP, which shows the expected monopolar texture. Note that the two WPs shown in Fig.~\ref{fig2}(a) show the same monopolar texture: They possess the same topological charge.

\begin{figure}[tbp]
    \centering
    \includegraphics[width=0.99\linewidth]{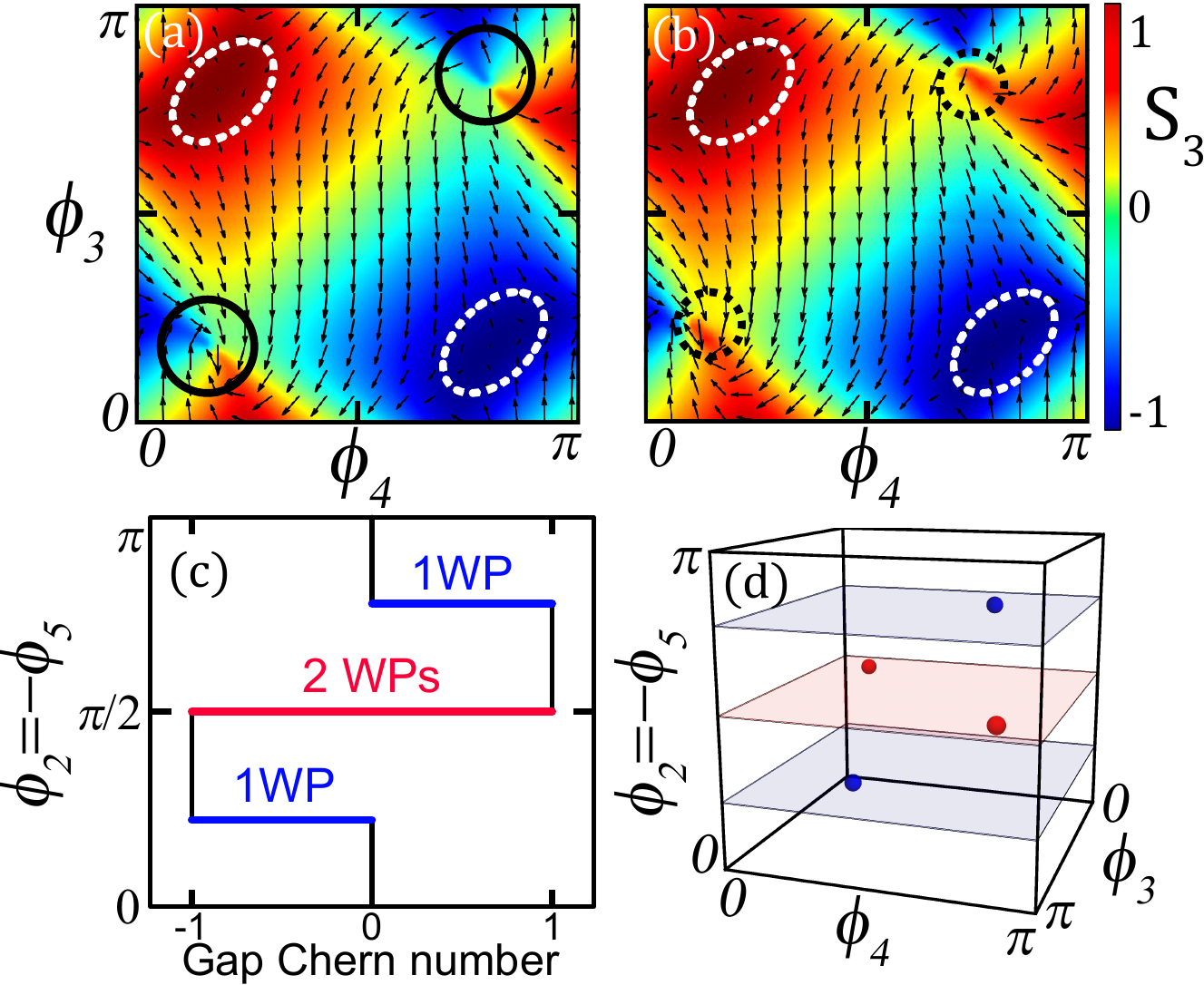}
    \caption{(a,b) $S_3$ pseudo-spin texture for $\phi_2=\pi/2$ (a) and $\phi_2=\pi/2-\pi/20$ (b). Arrows: $(S_2,S_1)$ pseudo-spin. Black (white) ellipses surround WPs along the diagonal (anti-diagonal) direction. Dashed ellipses surround gapped WPs. (e) Gap Chern number while varying the control phase $\phi_2=-\phi_5$. The black dashed lines indicate the position of the panels (a-b) in this figure. (f) WP positions in the 3D parameter space. Color denotes the sign of their topological charge (red - positive, blue - negative).}
    \label{fig3}
\end{figure}

The pentagonal geometry naturally breaks the inversion symmetry, allowing Weyl points existence. On the other hand, TR symmetry (which maps $\phi_i$ to $-\phi_i$) imposes that WPs appear in groups of 4~\cite{nielsen1981absence,riwar2016multi}. It means that there exists (at least) two other WPs in this 3D subspace.
Fig~\ref{fig3}(a) shows the pseudo-spin texture of the lower band at $\phi_2=\pi/2$. The black circles indicate the WPs. Furthermore, there are two maxima of circular polarization along the axis $\phi_3=-\phi_4$ (white ellipses), associated with a nonzero Berry curvature distribution~\cite{suppl}, due to the presence of the two additional WPs at $\phi_2\neq\pi/2$ that are gapped for the parameters shown.
%These maxima are associated to a non-zero Berry curvature distribution  as shown on Fig ~\ref{fig3}(c). They correspond to gapped Dirac points. 
Moving away from $\phi_2=-\phi_5=\pi/2$, the degeneracy disappears while a circular polarization appears (see Fig~\ref{fig3}(b)). The two other gapped WPs remain on the $\phi_3=-\phi_4$ axis.
Fig.~\ref{fig3}(c) shows the gap Chern number versus $\phi_2=-\phi_5$. Changes in the Chern number value are linked with gap closings at WPs. This occurs at $\phi_2=-\phi_5=\pi/2$ as shown before with the Chern number changing from -1 to +1 (because there are two WPs). The two other transitions occur at $\phi_2\approx \pm \pi/4.5$. They are each associated with a decrease by 1 of the Chern number because they correspond to a band crossing occurring trough a single (negative) WP.
Fig~\ref{fig3}(d) summarizes the previous results by showing the coordinates of the 4 WPs in the subspace ($\phi_2=-\phi_5,\phi_3,\phi_4$). 
WPs being robust against perturbations, moving away from the condition $\phi_2=-\phi_5$ only move them. We will now study the full 4D parameter space and observe their trajectories.

%%%%%%%%%%%%%%%%%%%%%%%%%%%%%%%%%%%%%%%%%%%%%%%%%%%%%%%%%%%%%%%
%%%%%%%%%%%%%%%%%%%%%%%%%%%%%%%%%%%%%%%%%%%%%%%%%%%%%%%%%%%%%%%
%%%%%%%%%%%%%%%%%%%%%%%%            %%%%%%%%%%%%%%%%%%%%%%%%%%%
%%%%%%%%%%%%%%%%%%%%%%%%%%%%%%%%%%%%%%%%%%%%%%%%%%%%%%%%%%%%%%%
%%%%%%%%%%%%%%%%%%%%%%%%%%%%%%%%%%%%%%%%%%%%%%%%%%%%%%%%%%%%%%%
\emph{4D parameter space.--}
\begin{figure}[tbp]
    \centering
    \includegraphics[width=0.99\linewidth]{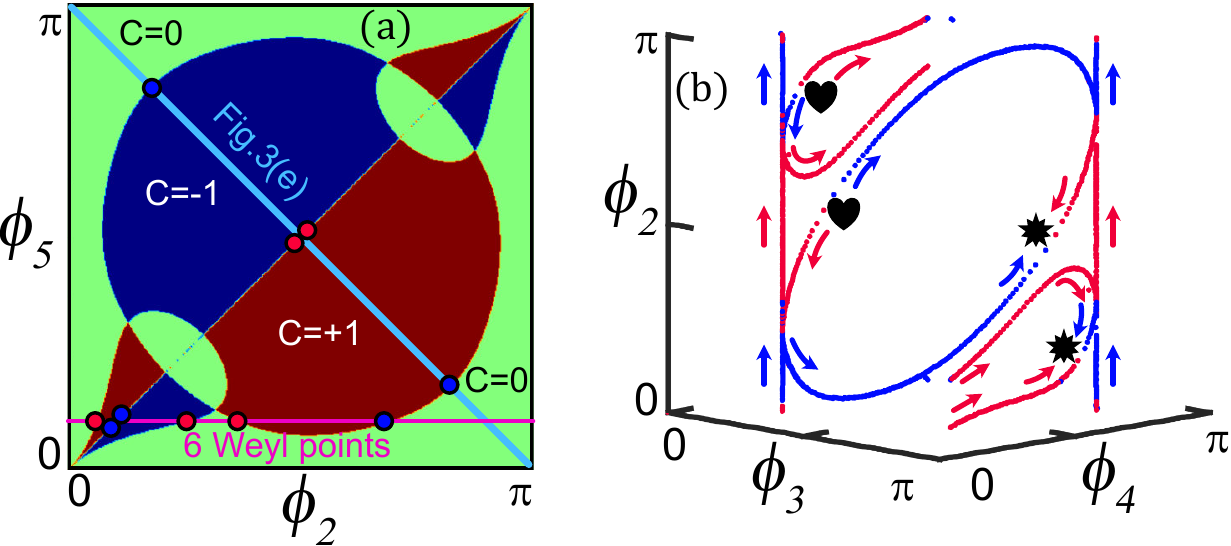}
    \caption{(a) Chern number of the lowest band in the $\left( \phi_3,\phi_4 \right)$ subspace for different $\phi_{2,5}$. The light blue line corresponds to $\phi_2=-\phi_5$ whereas the pink one represents a 3D subspace with broken TR symmetry and 6 WPs. (b) $\phi_{2,3,4}$ parameter space for $\phi_5\in[0;\pi[$. Arrows represent direction of the trajectories of WPs in parameter space. The hearts represent the birth of WPs and the stars their annihilation. Blue (red) points denote positive (negative) WPs.}
    \label{fig4}
\end{figure}
We go back to the full 4D parameter space $\phi_{2,3,4,5}$. To efficiently study this vast parameter space, we develop a model restricted to the two $\Tilde{p}$ states describing accurately the simulations shown previously. This model gives the 2-band 3D Hamiltonian:
\begin{equation}\label{ham}
    H=\mathbf{\Omega}\cdot \bm{\sigma},
\end{equation}
where $\bm{\sigma}$ is the 3-component vector of Pauli matrices and $\mathbf{\Omega}$ is constructed by interpreting the numerical results.
We first derive the in-plane terms $\Omega_{x,y}$. For two facing superfluids in 1D, the band is a cosine of twice their phase difference~\cite{septembre2021parametric}. Thus, looking at the modes profile in real space, we deduce which phases control them~\cite{suppl}. For instance, the energy of the $\Tilde{p}_x$ mode is given by two 1D bands with phase differences between $\phi_1$ and $\phi_{3,4}$ respectively. A similar reasoning for $\Tilde{p}_y$ leads to:
\begin{equation}
    \Omega_x=\left(\cos(2\phi_{2,5})+\cos(2\phi_{3,4})\right)/4-\cos(2\phi_{1,3})-\cos(2\phi_{1,4})
\end{equation}
and (anti-)diagonal states to:
\begin{equation}
    \Omega_y=\cos(2\phi_{2,4})-\cos(2\phi_{3,5}),
\end{equation}
where $\phi_{i,j}=\phi_i-\phi_j$. The $1/4$ factor appears because the pentagon orientation favors $\Tilde{p}_x$.
Then, studying the successive reflections of a wave starting from $\phi_1$ to $\phi_3$, we see that they lead to an overall rotating current~\cite{suppl}. This gives birth to (anti-)vortex states~\cite{suppl}, which are topological singularities in real space around which there is a winding of the argument of the wave function. The corresponding term is:
\begin{equation}
    \Omega_z=\sum_i\sin(2\phi_{i,i+2}).
\end{equation}
Altogether, we obtain the complete Hamiltonian of Eq.~\eqref{ham} which fits very well the numerical results shown previously~\cite{suppl}. This facilitates the exploration of the full 4D parameter space. In Fig.~\ref{fig4}(a) the 2D Chern number (for $\phi_{3,4}$ as in Eq.~\eqref{Chern}) is plotted for different values of $\phi_{2,5}$. We see that the Chern number takes nonzero values for wide areas in this space. We observe a very good agreement between the analytical and numerical results: The anti-diagonal (blue line) corresponding to the condition $\phi_2=-\phi_5$ behaves as expected. Beginning at $\phi_2=-\phi_5=0$ and increasing it, the system goes from a zero Chern number to $-1$ after meeting a negative WP, then $+1$ after meeting 2 positive WPs, then going back to the trivial case after a last WP. Note that the lines in 4D cannot be assigned a topological charge whereas points in 3D can.

Taking $\phi_5$ as a tuning parameter, we can explore the full 4D parameter subspace by plotting the $\phi_{2,3,4}$ space for each value of $\phi_5$, as in Fig.~\ref{fig4}(b) (varying $\phi_5$ with time gives a short movie~\cite{suppl}).
We can see the trajectories of the 4 WPs and the birth and annihilation of additional WPs. 
There can be 6 WPs (see Fig.~\ref{fig4}(a)), which is impossible if TR symmetry is preserved. 
Indeed, a WP at coordinates $(\phi_{2,3,4,5}^{(0)})$ has its TR twin at $(-\phi_{2,3,4,5}^{(0)})$, whereas when setting $\phi_5$ to a given value, the WP at $(\phi_{2,3,4}^{(0)},\phi_5)$ will have its twin at $(-\phi_{2,3,4}^{(0)},-\phi_5)$ and thus will not be met since $\phi_5\neq -\phi_5$. So, there exist 3D subspaces with broken TR symmetry leading to an even number of WPs (not necessarily multiples of 4).

\emph{Conclusion.--}
We investigate a polaritonic analog of multi-terminal Josephson junctions with 5 terminals. We show that it forms 4D synthetic bands with  topological 3D Weyl points. A topological transition can be probed by tuning the phases of pump lasers. The advantage of photonic systems is in the possibility of explicit experimental measurements of the eigenstates and thus of the band topology.
Further studies could explore higher dimensions and more exotic phases such as 3D Chern insulators~\cite{yang2019realization} and non-Hermitian degeneracies~\cite{zhen2015spawning,cerjan2019experimental,mc2020weyl}.

\begin{acknowledgments}
We acknowledge useful discussions with P. Kokhanchik.
This research was supported by the ANR Labex GaNext (ANR-11-LABX-0014), the ANR program "Investissements d'Avenir" through the IDEX-ISITE initiative 16-IDEX-0001 (CAP 20-25) and the European Union's Horizon 2020 program, through a FET Open research and innovation action under the grant agreement No. 964770 (TopoLight).
\end{acknowledgments}

\bibliography{biblio}

\renewcommand{\thefigure}{S\arabic{figure}}
\setcounter{figure}{0}
\renewcommand{\theequation}{S\arabic{equation}}
\setcounter{equation}{0}

\section{Supplemental material}

\subsection{Link between the laser and superfluid phases}
\begin{figure}[tbp]
    \centering
    \includegraphics[width=0.9\linewidth]{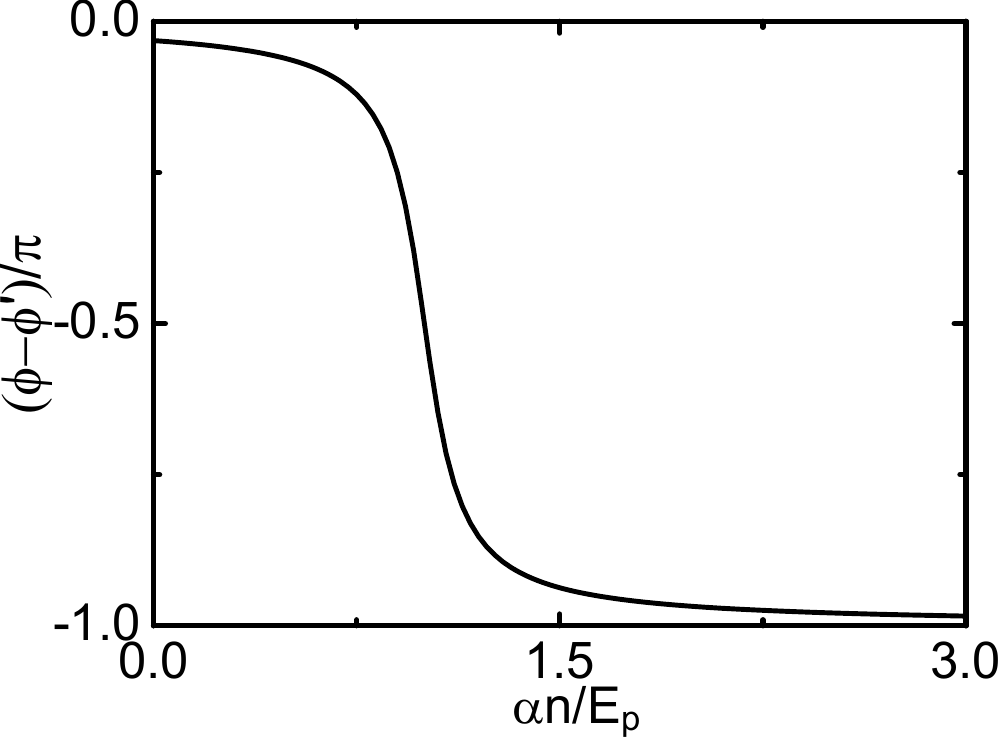}
    \caption{Phase of the superfluid with respect to the parameters. $\gamma=0.1\,E_p$.}
    \label{phases}
\end{figure}

In the main text, we define the phase of the $i$-th pump laser by $\phi'_i$ and the phase of the $i$-th superfluid by $\phi_i$, the latter being determined by the former. In this section, we provide details on the precise relation between those two phases.
The mathematical relation between the two phases is given by:
\begin{equation}
    \phi-\phi'=\arctan\frac{\gamma}{E_p+\alpha n}.
    \label{phaseEq}
\end{equation}
We can plot the difference of the two phases with respect to the ratio between the interactions and the pump energy, as shown in Fig.~\ref{phases}.

\subsection{Confined states in the 5-terminal Josephson junction}\label{SS1}
\begin{figure}[tbp]
    \centering
    \includegraphics[width=0.9\linewidth]{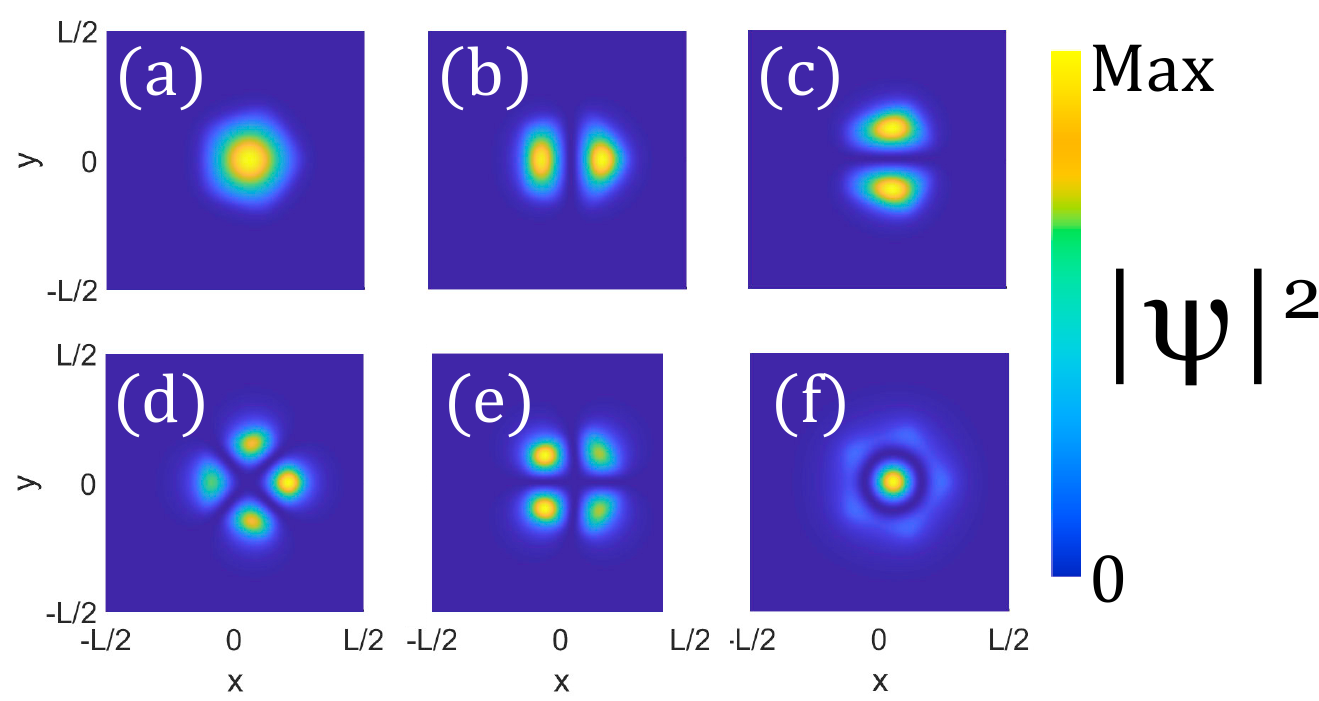}
    \caption{(a-f) Density probability of the $\Tilde{s}$ state~(a), $\Tilde{p}_x$ state~(b), $\Tilde{p}_y$ state~(c), $\Tilde{d}_{x^2-y^2}$ state~(d), $\Tilde{d}_{xy}$ state~(e) and $2\Tilde{s}$ state~(f). $L=12\,\mu$m for all panels.}
    \label{figS1}
\end{figure}

Solving the Bogoliubov-de Gennes equations without non-diagonal terms approximates a quantum well with no Andreev-like bound states. The states that come from the confinement brought by interactions are plotted in Fig.~\ref{figS1}.
Formally, any bound state wavefunction takes the form:
\begin{equation}
    \ket{\Psi}=\sum_{i=1}^\infty c_i \ket{\psi_i},
\end{equation}
where $\{\ket{\psi_i}\}$ forms a basis of states and $c_i=\braket{\Psi|\psi_i}$.
We assume that the first six states are sufficient to form an approximate basis for the states we observe for small traps, as done in our work. Thus, $\Psi$ is rather defined as:
\begin{widetext}
\begin{equation}
    \ket{\Psi}=c_s \ket{\psi_s}+c_{p_x} \ket{\psi_{p_x}}+c_{p_y} \ket{\psi_{p_y}}+c_{d_{x^2+y^2}} \ket{\psi_{d_{x^2+y^2}}}+c_{d_{xy}} \ket{\psi_{d_{xy}}}+c_{2s} \ket{\psi_{2s}},
\end{equation}
\end{widetext}
and we take $|c_s|^2+|c_{p_x}|^2+|c_{p_y}|^2+|c_{d_{x^2+y^2}}|^2+|c_{d_{xy}}|^2+|c_{2s}|^2=1$ which is in our study approximately true (up to the 5th decimal).

\subsection{Vortices, antivortices and winding number}
In the main text, the interplay between the phase of a bound states and the different phases imprinted by the superfluids is discussed. This leads, for certain phase patterns, to the presence of bound state vortices and anti-vortices, which are toplogical singularities characterized by non-zero winding number:
\begin{equation}\label{winding}
    w=\frac{1}{2\pi}\oint\arg \psi (\mathbf{r}) \mathbf{dr},
\end{equation}
where $\mathbf{dr}$ forms a close loop around the center of the phase singularity (abrupt change of $2\pi$) in the counter-clockwise direction.
\begin{figure}[tbp]
    \centering
    \includegraphics[width=0.9\linewidth]{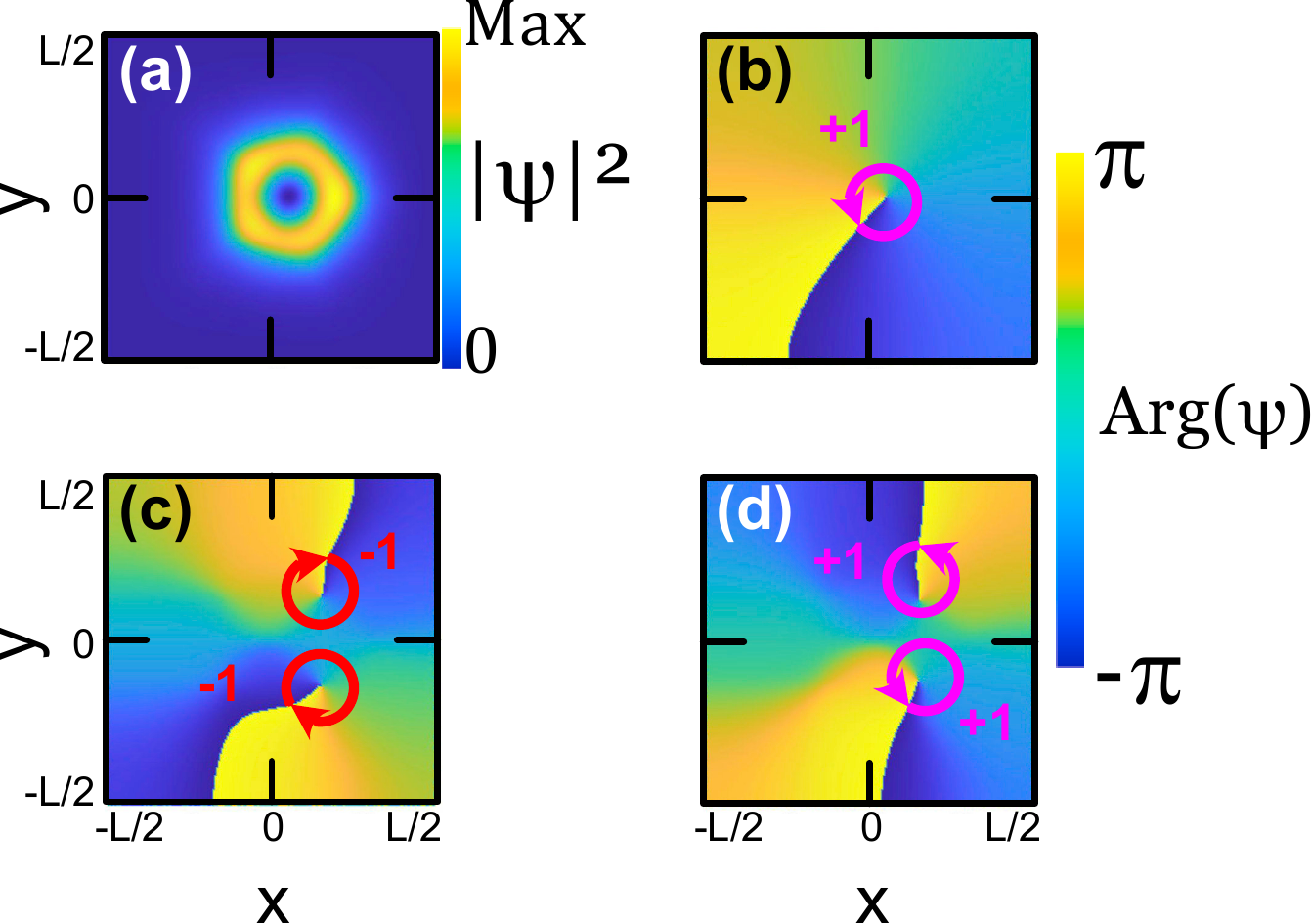}
    \caption{(a,b) Vortex state in the 5-terminal analog Josephson junction for $\phi_2=0$, $\phi_3=0$ and $\phi_4\approx\pi/4.5$. (a) Density probability distribution in real-space of a vortex state. The distribution for an anti-vortex is the same. (b) Phase of the wavefunction in real space. Note the positive phase winding around the central phase singularity. (c) Configuration with two anti-vortices, leading to a total winding number $w_{tot}=-2$. (d) Configuration with two vortices, leading to a total winding number $w_{tot}=+2$. The red arrows go clockwise (count for negative winding, see Eq.~\eqref{winding}) while the magenta ones go counter-clockwise (count for positive winding). For all panels, $a=1.7\,\mu$m and $L=12\,\mu$m.}
    \label{figS2}
\end{figure}
The phase pattern allows the formation of vortices and anti-vortices, as well as the existence of several (anti)-vortices at the same time, associated with non-trivial and eventually large total winding number, as illustrated by Fig.~\ref{figS2}.

\subsection{$\Tilde{s}$ band and its Andreev counterpart}
\begin{figure}[tbp]
    \centering
    \includegraphics[width=0.9\linewidth]{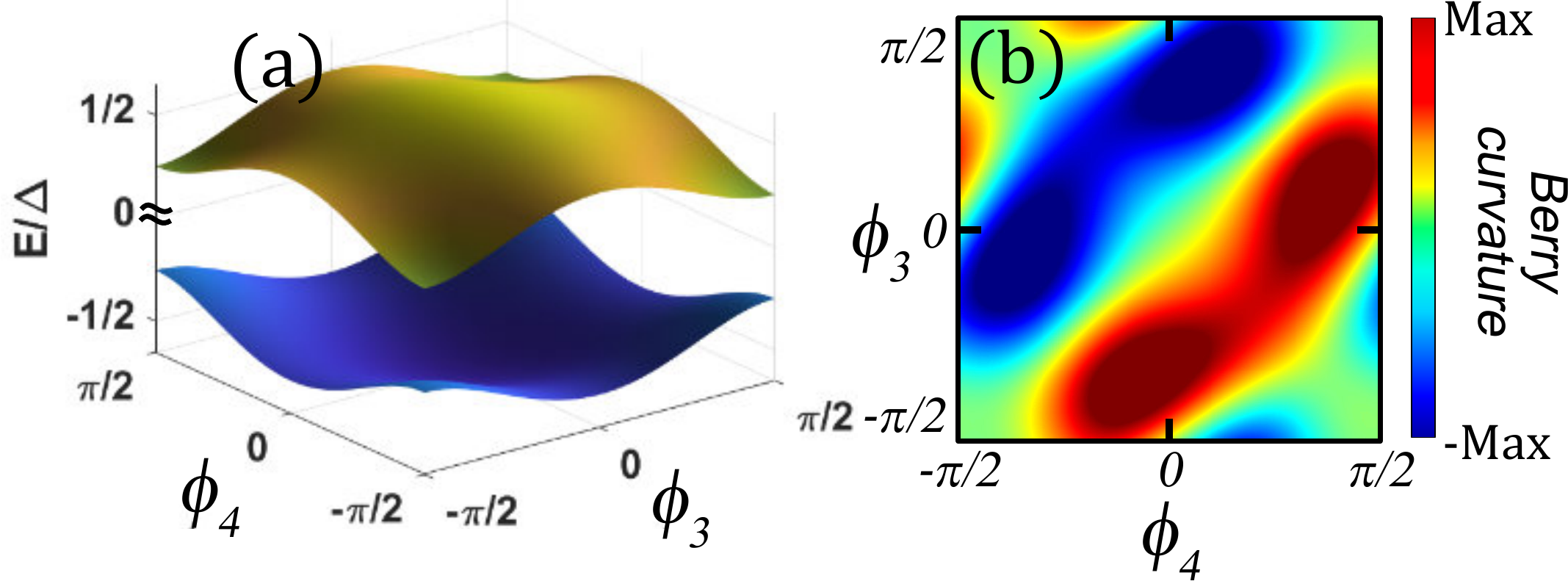}
    \caption{(a) $\Tilde{s}$ band (yellow) and its Andreev counterpart (blue) for $\phi_2=0$ and $a=1.2\,\mu$m. (b) Berry curvature distribution of the Andreev counterpart of the $\Tilde{s}$ band. Note that the positive and negative contributions are symmetric and compensate to give a trivial topology $\mathscr{C}=0$.}
    \label{figS3}
\end{figure}

In this section we will discuss the properties of the $\Tilde{s}$ band and of its Andreev counterpart. The bands are plotted in the $\left ( \phi_3,\phi_4 \right )$ subspace in Fig.~\ref{figS3}(a). 
In 1D~\cite{septembre2021parametric}, only the Andreev counterpart bands were considered to be topological because of a $\pi$ Zak phase whereas the other had a $0$ Zak phase. In 2D, we can still see that the geometry of the Andreev counterpart bands is non-trivial with a strong Berry curvature distribution (see Fig.~\ref{figS3}(b)). However, the contribution of positive and negative sign compensate each other, and the $\pi$ Zak phase in 1D does not translate into a nonzero Chern number in 2D in general.

\subsection{Berry curvature computation}
The Berry curvature distribution is calculated using the method duly described in Ref.~\cite{fukui2005chern}. This method requires to organize a basis of elementary states on which to decompose the solutions found. We used the six states found and described previously. Since we considered mainly short junctions with $a<2\,\mu$m, we encountered only low energy states. The contribution to the Berry curvature of the projection on states of higher energy than $\Tilde{s}$ and $\Tilde{p}$ ones were negligible for quantum confined states. For the Andreev counterpart of the $\Tilde{p}$ states, their contribution is higher. 

\subsection{Gap Chern number}
In general, for a system of $N$ well-separated bands, the $n$-th gap ($n \in [1;N[$) separating the $n$-th from the $(n+1)$-th band is characterized by a gap Chern number:
\begin{equation}
    \mathscr{C}_n^{gap}=\sum_{i=1}^{n}\mathscr{C}_i,
\end{equation}
where $\mathscr{C}_i$ refer to the Chern number of the $i$-th band.
In our system, there is an even number of bands which, half of them being below the pump detuning $E_p$ and half of them being above it. The Andreev counterpart of a band with Chern number $\mathscr{C}$ has a Chern number $-\mathscr{C}$. Moreover, for the two $\Tilde{p}$ bands, topological singularities when removed let appear a gap separating bands of opposite Chern number. Thus, we have $\mathscr{C}_{N/2}^{gap}=0$, and the Chern number of a gap separating the two $\Tilde{p}$ bands is always equal to the Chern number of the lower band.

\subsection{Pseudo-spin definition}
We define the pseudo-spin components as follows:
\begin{eqnarray}
    S_1 & = & \frac{|c_{p_x}|^2-|c_{p_y}|^2}{|c_{p_x}|^2+|c_{p_y}|^2},\\
    S_2 & = & \frac{|c_{p_x+ p_y}|^2-|c_{p_x-p_y}|^2}{|c_{p_x+ p_y}|^2+|c_{p_x-p_y}|^2},\\
    S_3 & = & \frac{|c_{p_x+i p_y}|^2-|c_{p_x-i p_y}|^2}{|c_{p_x+i p_y}|^2+|c_{p_x-i p_y}|^2},
\end{eqnarray}
where $c_i$ is the scalar product of the bound state with the state $\Tilde{i}$. For instance, a $\Tilde{p}_x$ state shows $|c_{p_x}|^2=1$ giving $S_1=1$ while an anti-vortex state $\Tilde{p}_x-i\Tilde{p}_y$ shows $|c_{p_x-i p_y}|^2=1$ and $S_3=-1$.

\subsection{Berry curvature near Weyl points}
\begin{figure}[tbp]
    \centering
    \includegraphics[width=0.99\linewidth]{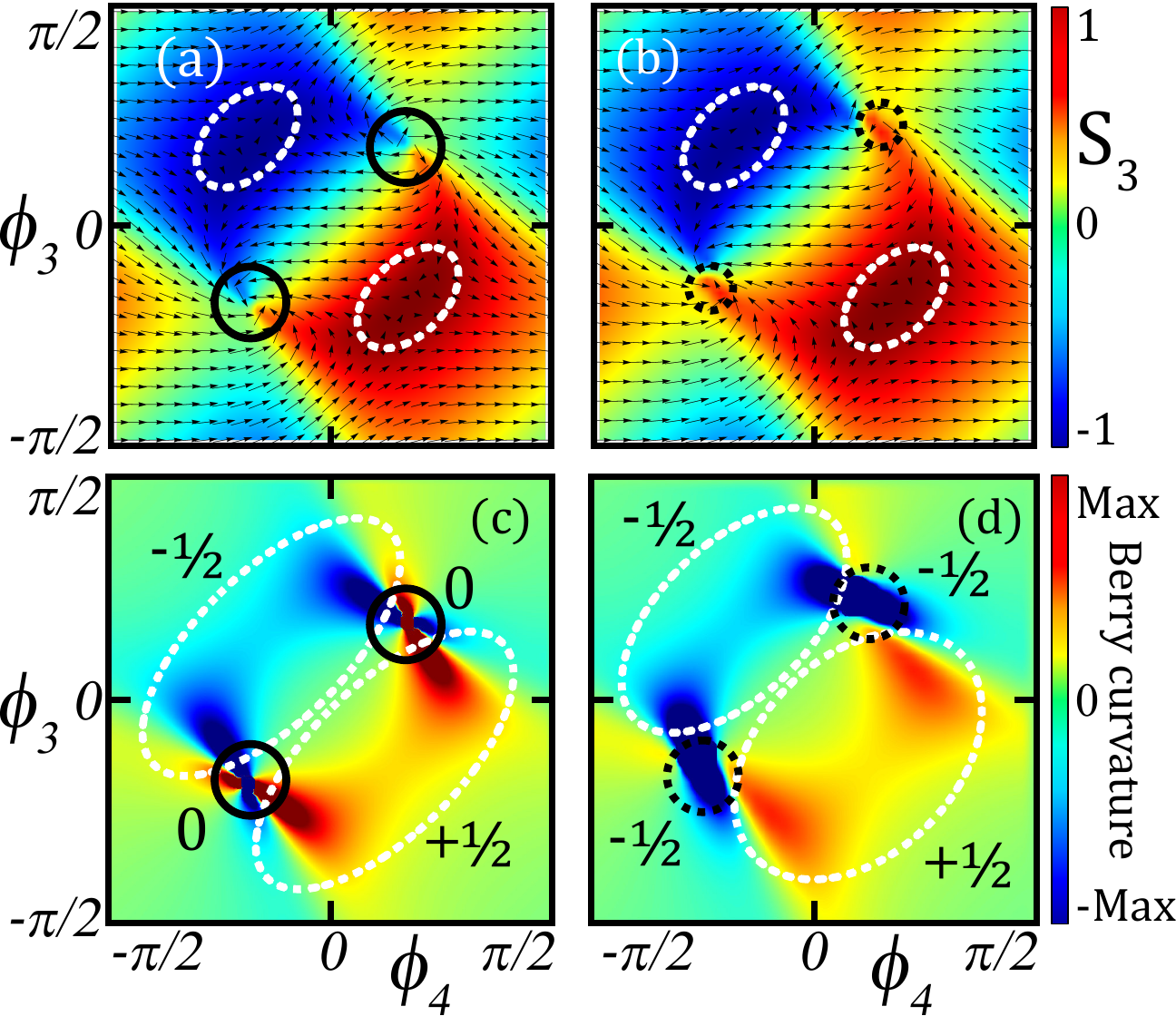}
    \caption{(a,b) $S_3$ pseudo-spin texture for $\phi_2=\pi/2$ (a) and $\phi_2=\pi/2-\pi/20$ (b). Arrows direction denote $(S_2,S_1)$ pseudo-spin. (c,d) Berry curvature distribution for $\phi_2=\pi/2$ (c) and $\phi_2=\pi/2-\pi/20$ (d). Black (resp. white) ellipses surround Weyl points along the diagonal (resp. anti-diagonal) direction. Dashed ellipses indicate that the point is gapped for this set of parameters.}
    \label{figS7}
\end{figure}
To calculate the Chern number, we compute the Berry curvature in the full 2D parameter space $\phi_{3,4}$ and integrate it. Fig~\ref{figS7} shows the $S_3$ component together with the Berry curvature at the position of two Weyl points and just next to it. We can clearly see the contributions of the four Weyl points in the Berry curvature when all Weyl points are gapped, which gives an overall nonzero Chern number.

\subsection{Explanations on the analytical model}
In the main text, we define the analytical model:
\begin{equation}\label{ham2}
    H=\mathbf{\Omega}\cdot \bm{\sigma}.
\end{equation}

\begin{figure}[tbp]
    \centering
    \includegraphics[width=0.99\linewidth]{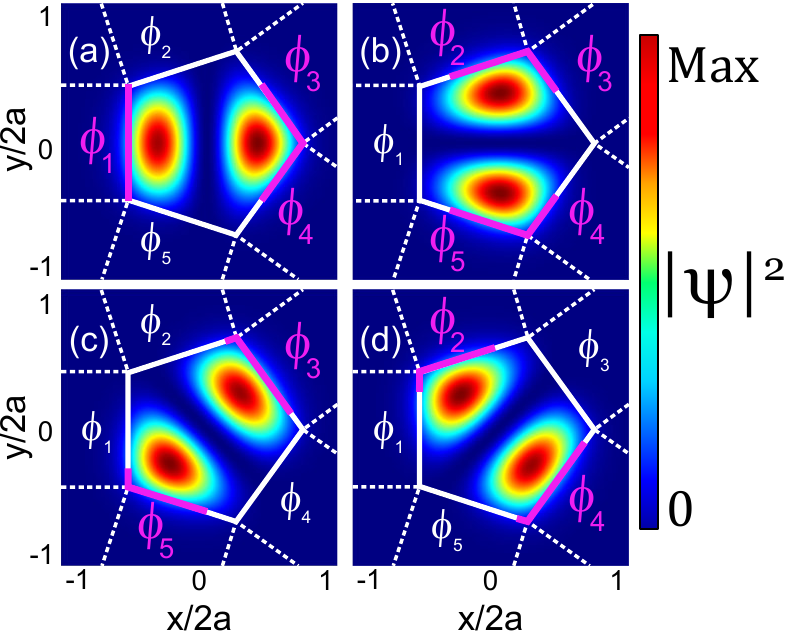}
    \caption{The interplay of the $\Tilde{p}_{x,y}$ and (anti-)diagonal states and the phases of the phases of the superfluids. (a) $\Tilde{p}_{x}$ state parametrized by $\phi_{1,3}$ and $\phi_{1,4}$. (b) $\Tilde{p}_{y}$ state parametrized by $\phi_{2,5}$ and $\phi_{3,4}$. (c) Diagonal state parametrized by $\phi_{3,5}$. (d) Anti-diagonal state parametrized by $\phi_{2,4}$.}
    \label{figS4}
\end{figure}

The in-plane contribution $\Omega_{x,y}$ can be deduced by the analysis of the shape of the $\Tilde{p}_{x,y}$ and (anti-)diagonal states. Considering Fig.~\ref{figS4}, we can clearly see that the different modes can be decomposed into 1D $p$ states with appropriate phases. We can then apply the result obtained in 1D~\cite{septembre2021parametric} that states form bands of cosine twice the phase difference between the two phases and find $\Omega_{x,y}$.

\begin{figure}[tbp]
    \centering
    \includegraphics[width=1.0\linewidth]{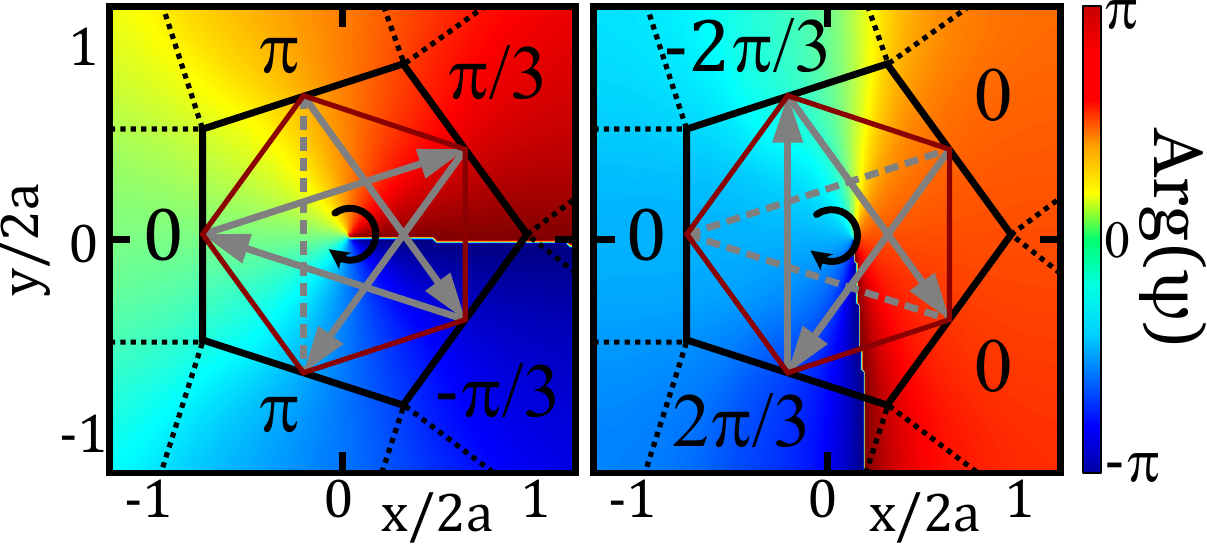}
    \caption{The formation of the non-zero angular momentum $p_x+i p_y$ superposition due to the Josephson currents giving rise to the $\sigma_z$ term opening the gaps at the Weyl points. (a) and (b) show 2 different configurations where the phase patterns imply the existence of a global rotation of the phase.}
    \label{figS5}
\end{figure}

The out-of-plane component is of different origin because it comes from the Josephson current. However, we will again invoke 1D results. The path of a particle going from the face 1 to the face 4 is plotted in Fig.~\ref{figS5}. The particle is then reflected by both normal and Andreev reflections through the coefficients~\cite{septembre2021parametric}:
\begin{equation}\label{refl-coeffs}
  \begin{array}{r c l}
      r_N & = & \frac{(k_I-i\kappa_-)(k_A+i\kappa_+)|u|^2 + (k_A+i\kappa_-)(k_I-i\kappa_+)|v|^2}{ (k_I+i\kappa_-)(k_A+i\kappa_+)|u|^2 + (k_A+i\kappa_-)(k_I+i\kappa_+)|v|^2},\\
    r_A & = & \frac{2 i \sqrt{k_I k_A} (\kappa_+-\kappa_-) |u||v| }{ (k_I+i\kappa_-)(k_A+i\kappa_+)|u|^2 + (k_A+i\kappa_-)(k_I+i\kappa_+)|v|^2}e^{-2i\phi}.
  \end{array}
\end{equation}
For certain phase configurations, the successive reflections lead to an overall nonzero current due to the Josephson current which is a sine of twice the phase difference between the two facing phases~\cite{septembre2021parametric}. This current is at the origin of the vorticity of states associated with topology, and explains the overall topology of the system.

\subsection{Comparison between analytics and numerics}
\begin{figure}[tbp]
    \centering
    \includegraphics[width=1.0\linewidth]{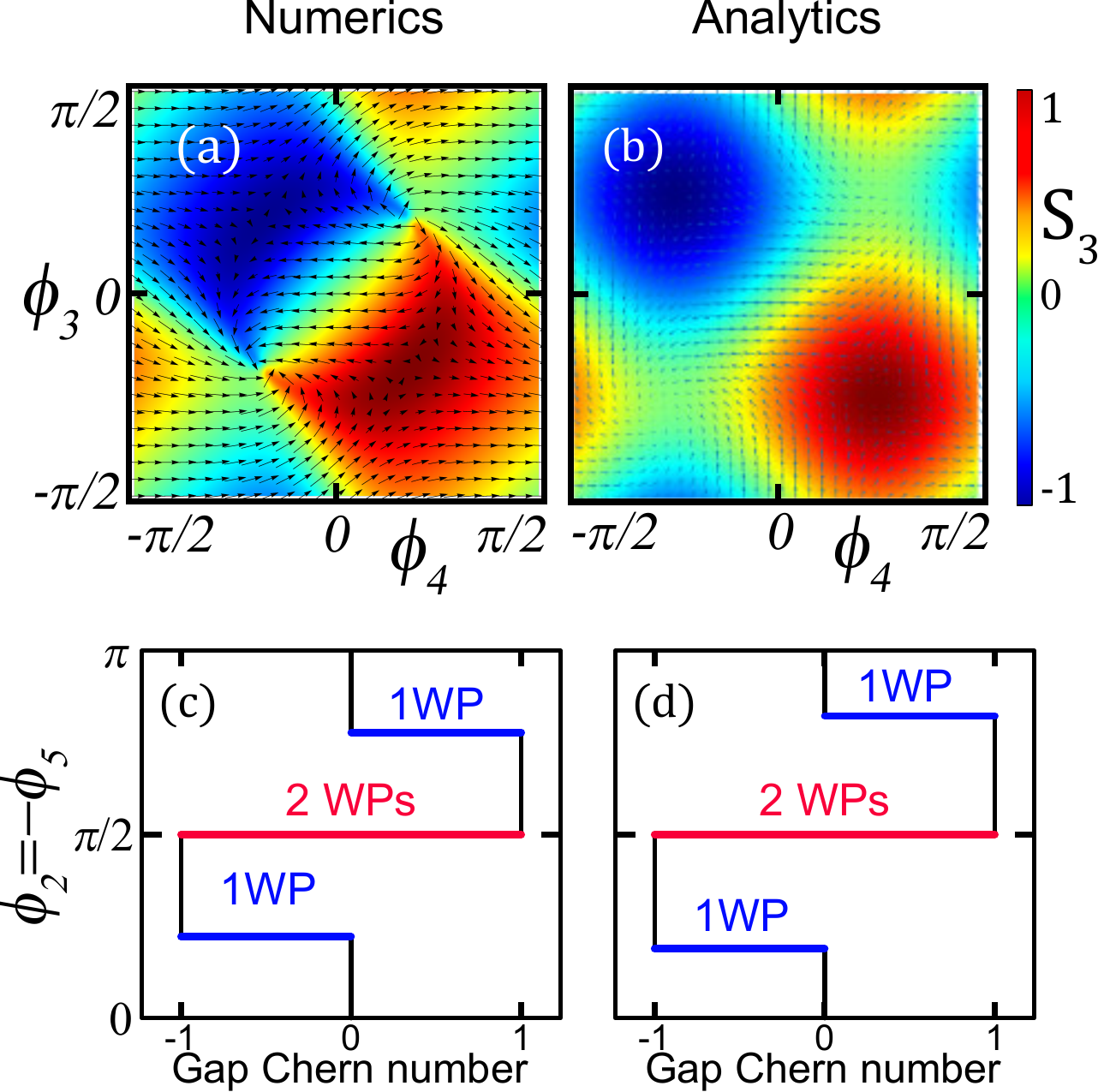}
    \caption{(a,b) Third pseudo-spin component $S_3$ for $\phi_2=-\phi_5=\pi/2$ calculated from numerical simulations (a) and our analytical model~\eqref{ham} (b). The direction of the arrows indicates the $(S_2,S_1)$ pseudo-spin. (c,d) Gap Chern number with respect to $\phi_2=-\phi_5$ calculated from numerical simulations (c) and our analytical model~\eqref{ham} (d). Note the excellent agreement.}
    \label{figS6}
\end{figure}

In the main text, a first part is devoted to the numerical study of a peculiar 3D subspace of the full 4D parameter space where $\phi_2=-\phi_5$. As already stated in the main text, this is more comfortable to study numerically the degeneracies because the 2D space where we plot the bands ($\phi_{3,4}$) is then much more symmetric. Furthermore, this helps us to understand how the modes behave and to construct an analytical model that facilitates the exploration of the full 4D parameter space.
Fig.~\ref{figS6}(a,b) shows the $S_3$ pseudo-spin textures together with $S_{1,2}$ computed with numerical solving of Bogoliubov-de Gennes equations (a) and our analytical model (b). We clearly see the good agreement between the two. More precisely, the pseudo-spin behavior in the vicinity of the 4 Weyl points corresponds to what is observed in numerical simulations. This leads to a very good agreement in the calculations of the Berry curvature and subsequent Chern numbers for different values of $\phi_2=-\phi_5$ (see Fig.~\ref{figS6}(c,d)).

\subsection{Movie}
In the main text, we show the Weyl points trajectories and annihilation in the 3D parameter space $\phi_{2,3,4}$ in Fig.4(b). We provide here the link to a movie corresponding to this panel. The axes are the same, and the phase $\phi_5$ is varied with time. Blue/red circles denote positive/negative Weyl points. 

https://www.youtube.com/watch?v=jZQvpuWw6zo

In the movie, we can clearly see the motion of Weyl points in the 4D space, including the creation and annihilation of Weyl points.

\end{document}